\documentclass[prl,floats,twocolumn,aps,superscriptaddress,showpacs,floatfix]{revtex4}

\usepackage{amsmath}
\usepackage{graphicx}
\usepackage{color}

\newcommand{\be}{\begin{equation}}
\newcommand{\ee}{\end{equation}}
\newcommand{\bea}{\begin{eqnarray}}
\newcommand{\eea}{\end{eqnarray}}
\newcommand{\barr}{\begin{array}}
\newcommand{\earr}{\end{array}}

\begin{document}

\title{A Granular Brownian Ratchet Model}

\date{\today}

\author{Giulio Costantini}
\affiliation{Universit\`a di Camerino, Dipartimento di Fisica,
Via Madonna delle Carceri, I-62032 Camerino, Italy}

\author{Umberto Marini Bettolo Marconi}
\affiliation{Universit\`a di Camerino, Dipartimento di Fisica,
Via Madonna delle Carceri, I-62032 Camerino, Italy}

\author{Andrea Puglisi} \affiliation{Universit\`a di Roma ``La
Sapienza'', Dipartimento di Fisica, p.le Aldo Moro 2, I-00185 Roma,
Italy}

\begin{abstract}
We show by numerical simulations that a non rotationally 
symmetric body, whose orientation
is fixed and whose center of mass can only slide along
a rectilinear guide, under the effect of 
inelastic collisions with a surrounding gas of particles, displays
directed motion. We present a theory which explains how the lack of time
reversal induced by the inelasticity of collisions can be exploited
to generate a steady average drift. In the limit of an heavy 
ratchet, we derive an effective Langevin equation whose parameters
depend on the microscopic properties of the system and
obtain a fairly good quantitative agreement between the 
theoretical predictions
and simulations concerning mobility, diffusivity and average velocity.
\end{abstract}

\maketitle
The striking
contrast between the simplicity of models or experimental setups
and the richness and complexity of the observed phenomena
has contributed to generate much interest toward the Physics of Granular 
Media over the past two decades~\cite{intro0}.
A series of reasons, such as the
macroscopic nature of grains, their inelastic
collisions and the lack of true thermodynamic equilibrium
represent an obstacle to a straightforward
application of standard methods of statistical mechanics. 
Dilute granular systems, the
so-called granular gases~\cite{intro1}, today are a
privileged theoretical and experimental benchmark to test the
fundaments of kinetic theory and of non-equilibrium statistical
mechanics in general~\cite{intro2}.

This paper is inspired to a recent  numerical
experiment~\cite{vdb} where a Brownian ratchet, i.e. 
a mechanical device able to rectify thermal fluctuations~\cite{rat},
is obtained in a non-equilibrium system with energy conserving dynamics.
As shown by Van den Broeck, this rectification can be obtained
by coupling the ratchet to two thermal reservoirs at different temperatures
without violating the Second Principle of Thermodynamics.
We underline that in order to generate a Brownian ratchet,
two symmetries must be broken: the time reversal symmetry
(detailed balance) and rotational invariance of the object.

We depart from this work by proposing an even simpler device, the
granular ratchet, which contains the minimal ingredients necessary to
obtain directed motion. It is designed to achieve a non-equilibrium
stationary regime using the inelasticity and the consequent lack
of detailed balance~\cite{ap} together with the broken rotational
symmetry to extract work from a single source.

\begin{figure}[htbp]
\includegraphics[width=7cm,clip=true]{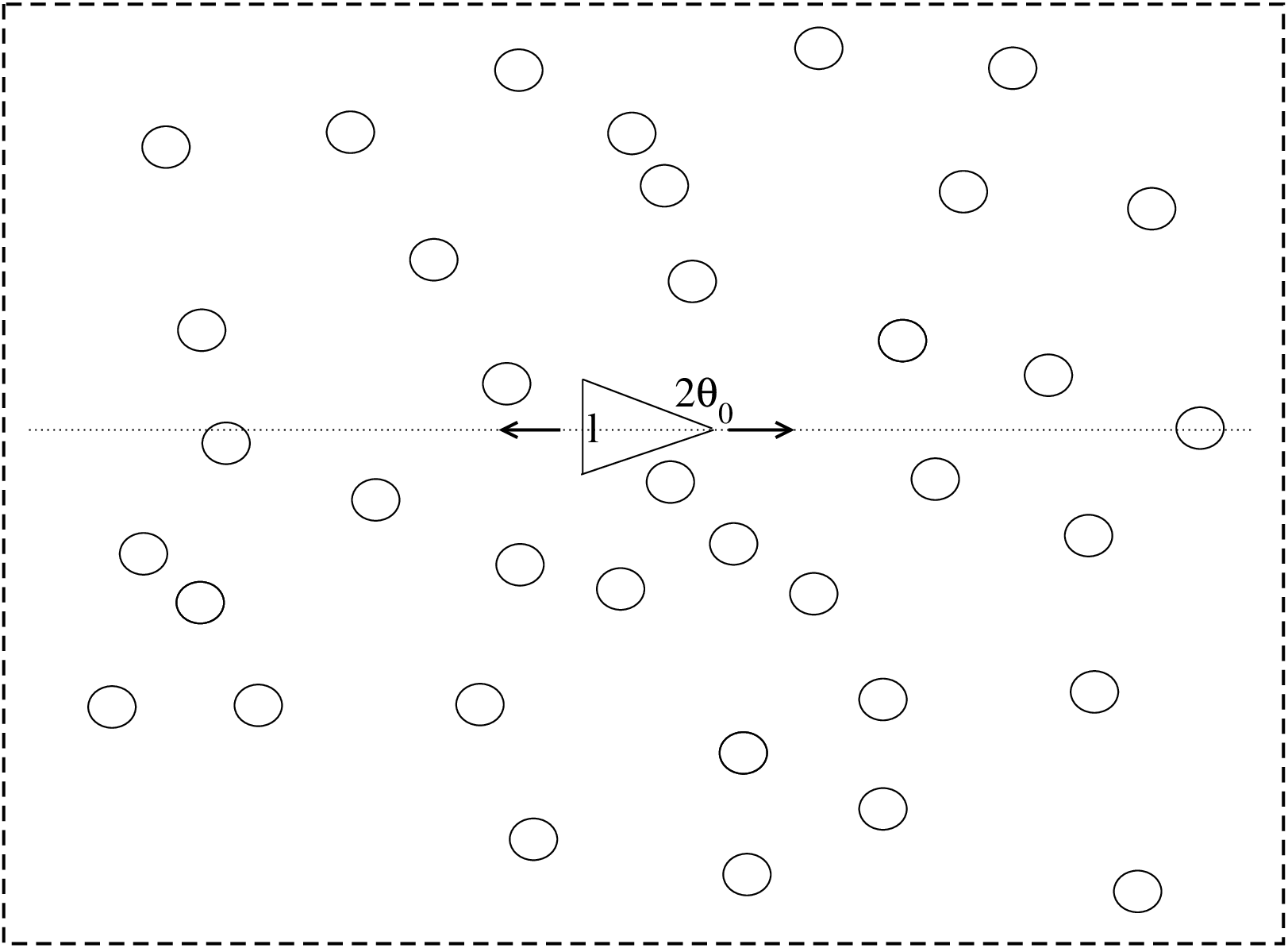}
\caption{Sketch of the 2D model. The triangle is constrained to move
only in the $\hat{x}$ (left/right) direction, while its orientation is
fixed, i.e. it cannot rotate. Gas particles collide against it and
occasionally receive energy from an external bath, in the form of
uncorrelated isotropic random kicks.
\label{fig:model}}
\end{figure}
The granular ratchet model, sketched in  Fig.~\ref{fig:model}, consists of a
triangular particle (the ratchet) of mass $M$, shaped as 
an isosceles triangle with base $l$
and angle opposite to the base $2 \theta_0$ and surrounded by 
a gas of $N$  disks of diameter $\sigma=1$ and mass $m=1$.
The ratchet can only slide, without
rotating, along the direction $x$, perpendicular to its base
and the whole system is enclosed in a squared box of side $L$
with periodic boundary conditions.
The $N+1$ particles undergo binary instantaneous collisions
described by the rule:
\be 
\label{eq:rule}
\mathbf{v}_i=\mathbf{v}'_i-(1+\alpha_{ij})c_{ij}
[(\mathbf{v}'_i-\mathbf{v}'_j)\cdot \hat{\mathbf{n}}]\hat{\mathbf{n}},
\ee where $\mathbf{v}$ and $\mathbf{v}'$ are the post-collisional and
pre-collisional velocities respectively, 
$\alpha_{ij}\le 1$ is the coefficient of restitution for that
particular collision, taking value $\alpha_d$ if both objects are
disks or value $\alpha_r$ if the ratchet is involved,
$\hat{\mathbf{n}}$ is the outward-pointing unit vector normal, in the
contact point, to the surface of particle $i$,
and $c_{ij}$ is a coefficient which takes the value $1/2$ if the
objects are both disks, or the value
$1/(1+\epsilon^2\hat{n}_x^2)$ if $j$ is the triangle, or the
value $\epsilon^2/(1+\epsilon^2\hat{n}_x^2)$ if $i$ is the
triangle, where $\epsilon^2=m/M$. Because of the constraint the vertical velocity
of the ratchet is always $0$.
The collision rule~\eqref{eq:rule} conserves the total momentum if $i$
and $j$ are disks, but conserves only the $x$-component of the momentum, 
when the triangle is involved. 
If $\alpha_{ij}=1$ the total kinetic energy is
also conserved. Three possible cases may be considered: (i) a pure
elastic gas where $\alpha_d=\alpha_r=1$, (ii) a mixed gas where
$\alpha_d=1$ and $\alpha_r<1$, (iii) a pure inelastic gas where
$\alpha_d<1$ and $\alpha_r<1$. In both cases (ii) and (iii) an
external driving mechanism is needed to attain a stationary state and
avoid indefinite cooling of the system. Here, we use a homogeneous random 
driving because it is the simplest and most studied in theoretical
literature~\cite{kicks_noi,kicks_altri}. In particular, our simulation
implements the non viscous version of this thermostat: all disks
receive, at a constant frequency, independent random Gaussian accelerations
with zero average and $T_b=1$
variance. In experiments one may easily reproduce such a
thermostat by placing the grains upon a horizontal plate vibrating at
a high frequency~\cite{kicks_exp}.
Since the results are qualitatively the same, we will
reduce the number of free parameters and restrict our simulations to
cases (i) and (ii), that is keep $\alpha_d=1$ and vary the
inelasticity of the triangle $\alpha_r$ only.  The system is simulated
by means of an event driven Molecular Dynamics algorithm. The triangle
is ``smoothed'' by approximating its three vertexes with arcs of
circle, with the condition of tangent continuity along the perimeter.

The gas is initialized (at $t=0$) by assigning to the disks 
non-overlapping random positions and Maxwellian velocities  
with zero average and unitary variance.  The
system forgets its initial configuration and attains a
stationary state. In the pure elastic case the total energy
$E=\frac{1}{2}\sum_i m_i v_i^2$ (where $m_i=m$ if $i$ is a disk and
$m_i=M$ if $i$ is the triangle) is strictly conserved, while in the
inelastic case, due to the action of the thermostat, it reaches a
stationary value depending on all the control parameters (frequency of the
thermostat, collision frequency, coefficients of restitution, masses,
etc.). Numerical simulations indicate that the probability
distribution function (pdf) for the velocity of gas particles and of
the ratchet are close to a Maxwellian. In the following we will
indicate as $T_g$ the stationary values of the gas temperature and $T_r$ the
ratchet temperature. All MD results given hereafter
are obtained using $N=1000$, $L=500$ (i.e. covered volume fraction
$\sim 4 \cdot 10^{-3}$), $\theta_0=\pi/6$, $l=10$, giving random
uncorrelated kicks to all gas particles every $2000$ seconds (gas
collision frequency is observed to be of the order of $1$ collision
every $20$ seconds).

\begin{figure}[htbp]
\includegraphics[angle=0,width=8cm,clip=true]{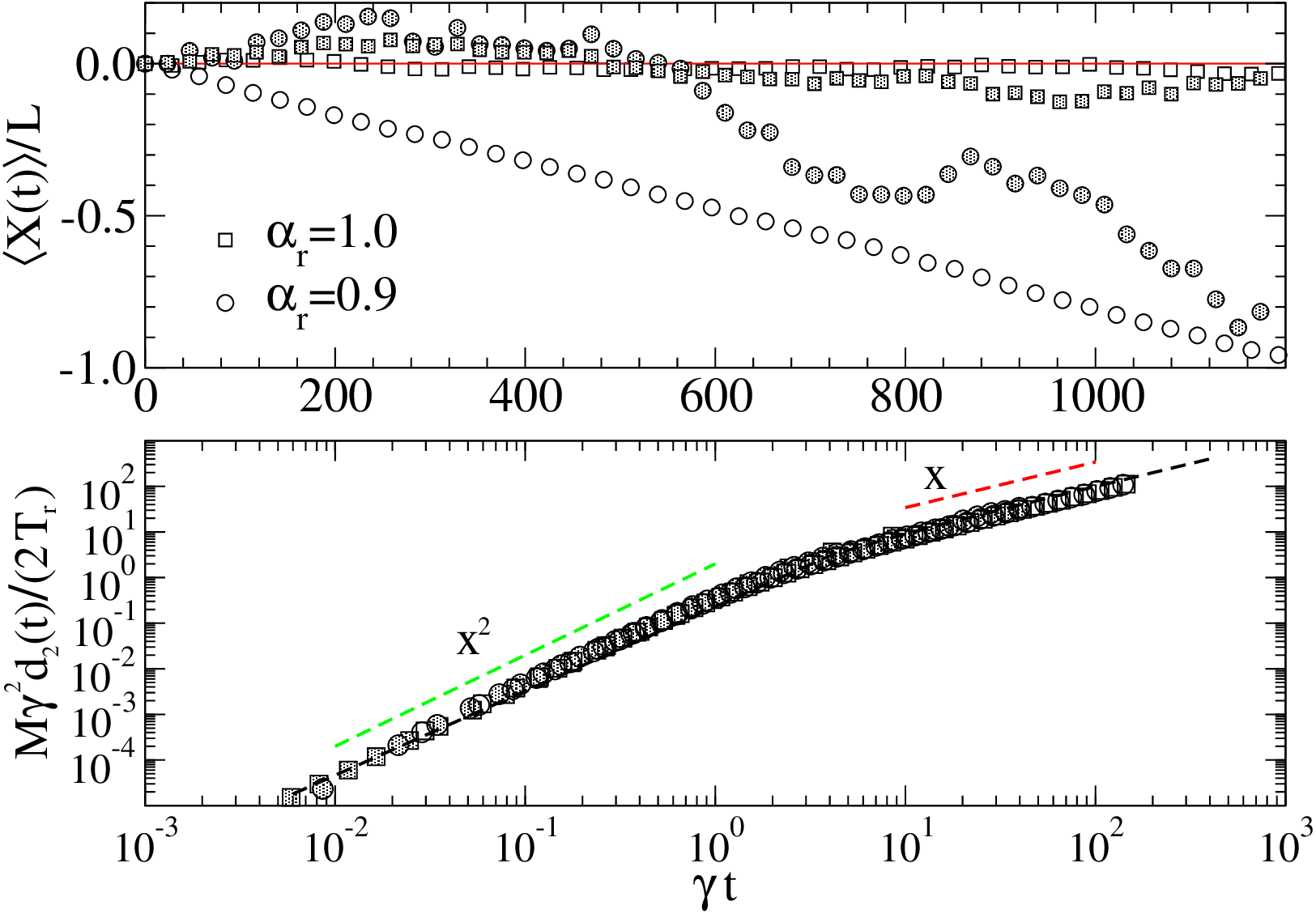}
\caption{Top frame: averaged trajectory of the tracer in MD (full
symbols) and DSMC (empty symbols) with $M=10$. Bottom frame: rescaled
mean squared displacement $d_2(t)$ (see text) for the same choices of
parameters. The power law $\sim x$ and $\sim x^2$ are drawn for reader's convenience
(dotted straight lines).
\label{fig:traj}}
\end{figure}

We focus here on the statistical behavior of the ratchet, whose
position and velocity at time $t$ are denoted as $X(t)$ and $V(t)$
respectively. Trajectories are averaged over $1000$ realizations
starting with different random configurations and discarding the
initial transient. Averaged trajectories for two particular choices of
the parameters are displayed in the top frame of Fig.~\ref{fig:traj},
showing that when the system is totally elastic no average motion
occurs for the triangle.  On the contrary, when inelasticity is
switched on, i.e. $\alpha_r<1$, even if the external driving mechanism
acts through random isotropic accelerations and without any privileged
direction, the triangle drifts with average velocity $\langle V
\rangle \neq 0$. In all our MD simulations we always observed a
negative velocity: the triangle on average moves toward its
base. We also studied the tracer self-diffusion, measuring the
quantity $d_2(t)=\langle (d(t)-\langle d(t) \rangle)^2 \rangle$ with
$d(t)=X(t)-X(0)$ the displacement of the tracer with respect to a
starting time $t=0$, taken when the whole system has become
stationary. This measure is presented in the bottom frame of
Fig.~\ref{fig:traj}, rescaled following the theory discussed
below. The usual Brownian behavior with a ballistic first stage and
diffusive asymptotics is observed. 

\begin{figure}[htbp]
\includegraphics[width=8cm,clip=true]{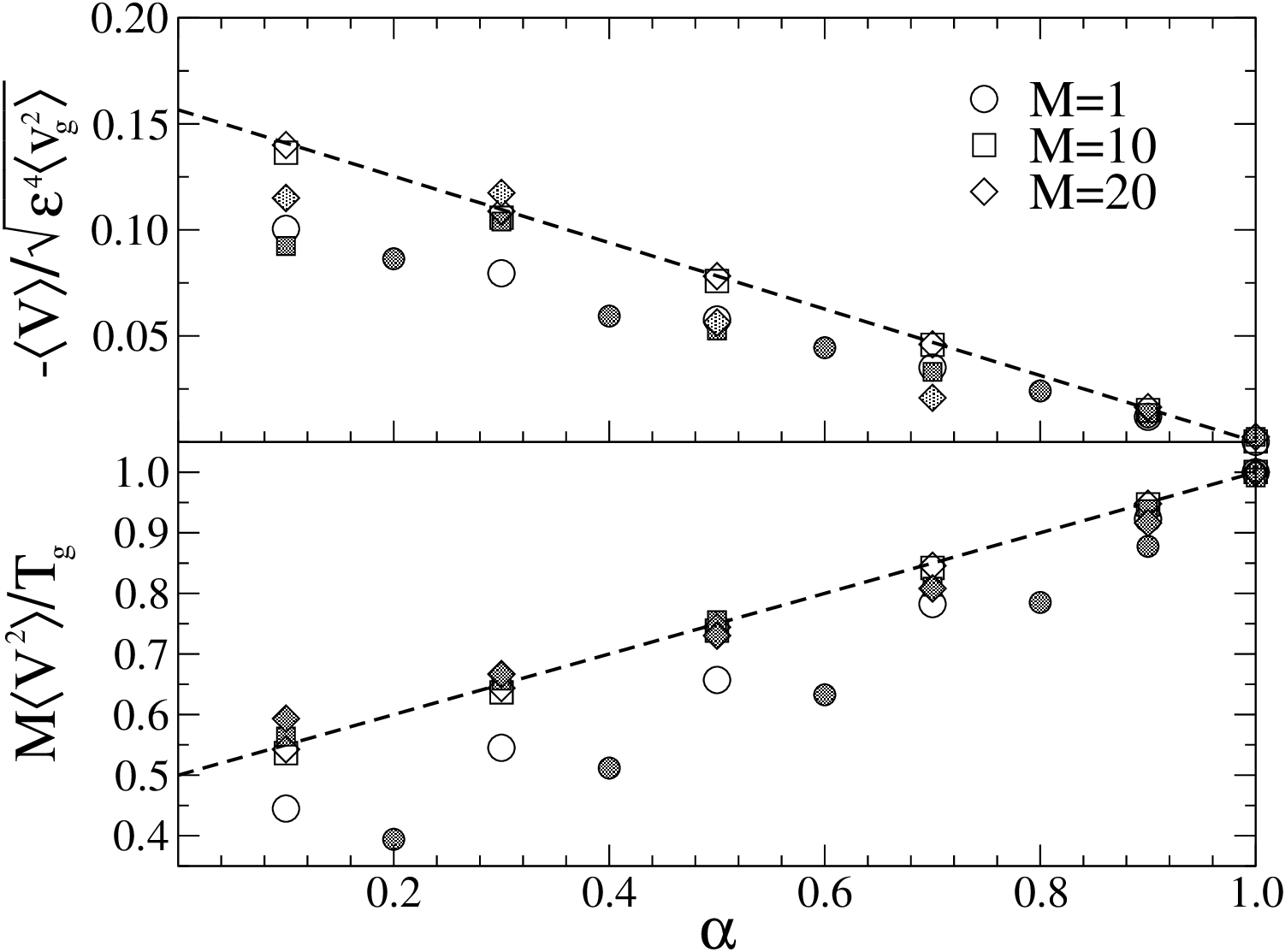}
\caption{Rescaled average velocities and energies of the tracer, both
from MD (full symbols) and DSMC (empty symbols) simulations. Averages
have been obtained with $10^3$ MD dynamics and $10^4$ DSMC
dynamics. Gas particles have always $m=1$. In MD,
$T_g=m \langle v_g^2 \rangle/2$ changes when changing the parameters ($\alpha$ and
$M$), in DSMC it is always $T_g=1$. 
\label{fig:all}}
\end{figure}
In Fig.~\ref{fig:all} the average velocities and
energies of the triangle, measured in MD (gray
symbols), are shown.
The absolute value of the average velocity of the tracer
increases as elasticity is reduced, showing the origin
of the effect. As the nice collapse suggests, the ratchet velocity is
proportional to the thermal velocity of the gas $\sqrt{2T_g/m}$ and to
$1/M$. Therefore, its measure becomes very difficult in the limit of
large $M$, being blurred by thermal noise. 
The bottom frame of Fig.~\ref{fig:all} illustrates the decrease of the
average ratchet energy, with respect to $T_g$, with increasing
inelasticity. For smaller ratchet masses the effect is stronger. We
have also considered different values of $\theta_0$ (not shown in
figure): it appears that reducing this angle is a way to increase the
ratchet effect, i.e. $|\langle V \rangle|$ increases. This is
consistent with the fact that non-zero motion of the triangle has
origin in its asymmetry. Our
next step is obtaining some analytical predictions to be compared with
numerical observation.

In the dilute gas limit, it is reasonable to study the ratchet dynamics
by means of a linearized Boltzmann equation
for its velocity pdf, $P(V,t)$,  which can be written as a 
Master Equation (ME) for a Markov process~\cite{vdb}:
\be \label{eq:boltz}
\frac{\partial P(V,t)}{\partial t}=\int
dV'~[W(V|V')P(V',t)-W(V'|V)P(V,t)] 
\ee 
where the transition rate is:
\bea
W(V|V')=\int_0^{2\pi}d\theta S F(\theta)\int_{-\infty}^{\infty}dv'_x
\int_{-\infty}^{\infty}dv'_y \rho \phi(v'_x,v'_y)  \nonumber \\
(\vec{V}'-\vec{v}')\cdot \hat{\mathbf{n}}\Theta
\big[(\vec{V}'-\vec{v}')\cdot \hat{\mathbf{n}} \big] \cdot
\delta[V-V_{post}(V',\vec{v}',\alpha_r,\epsilon)]  
\label{boleq}
\eea 
with $\rho$
the density of the gas, $V_{post}$ the
post-collisional ratchet velocity (see eq.~(\ref{eq:rule})), $\Theta$
the Heaviside step function, $S$ the perimeter length,
$\hat{\mathbf{n}}=(\sin\theta,-\cos\theta)$ and for the triangle
$S F(\theta)=\frac{l}{2\sin\theta_0}\big\{2\sin \theta_0
\delta(\theta-3\pi/2)+\delta(\theta -\theta_0)+
\delta[\theta-(\pi-\theta_0)]\big\}$. Following numerical evidence 
we approximate the velocity pdf of the gas,
$\phi(\mathbf{v})$, by a Maxwellian with zero mean and variance $T_g$.
It is straightforward to verify that detailed
balance, in the form $P(V)W(V'|V)=P(-V')W(-V|-V')$, holds  
if $\phi(\mathbf{v})$ is Gaussian and $\alpha_r=1$.

In order to gain a deeper insight it is convenient to approximate the
ME by a Fokker-Planck equation (FPE), 
from which we can extract the
analytical expression of the drift and diffusion terms.   
This is achieved by expressing the r.h.s. of eq. (\ref{eq:boltz})  by means of
the Kramers-Moyal (KM) expansion 
\be
\frac{\partial P(V,t)}{\partial
t}=\sum_{n=1}^\infty \frac{(-1)^n}{n!}\left(\frac{d}{dV}\right)^n
[j_n(V) P(V,t)]
\ee
where $j_n(V)=\int dV'(V'-V)^n W(V'|V)$.
By retaining only the first two terms we obtain the sought FPE, which 
can be still simplified by expanding these terms in the  
small parameter $\epsilon$.
The resulting expressions suggest a simple
physical picture, which can be illustrated with the help of the Langevin 
equation associated with the FPE:
\be  
\label{eq:langevin}
\dot{V}(t)=-\gamma V(t)+\frac{F}{M}+\Gamma(t)
\ee
The quantities $\gamma,F,\Gamma(t)$ are effective parameters and are related
to the original parameters by 
\be
\gamma=4\eta\rho l \epsilon\sqrt{\frac{T_g}{2\pi M}}(1+\sin \theta_0) 
\label{eq:visc}\\
\ee
\be
\frac{F}{M}=-
\rho l\frac{T_g}{M}\epsilon^2(1-\sin^2 \theta_0)\eta(1-\eta)
\label{eq:force}\\
\ee
\be
\langle \Gamma(t) \Gamma(t') \rangle=\frac{2 \gamma T_r}{M}
\delta(t-t') \;\;\;\;\;\;\;\; \langle \Gamma(t)\rangle=0  
\label{eq:fdt}\\
\ee
\be
1-\eta=1-\frac{T_r}{T_g}=\frac{1-\alpha_r}{2} 
\label{eq:temperatura}\\
\ee
Hence, for $\alpha_r<1$ the ratchet drifts with an average
negative velocity $\langle V(t)\rangle=F/ (M\gamma)$.
Indeed, the net velocity vanishes
linearly with $\epsilon\to 0$ and is very tiny for 
massive ratchets.
It is of interest to observe that in virtue of
eq. (\ref{eq:temperatura}) the net driving force is proportional
to the temperature difference $T_g-T_r$, so that
the tracer and the gas temperatures play role analogous
the two reservoir temperatures of the Brownian ratchet model.

To corroborate the robustness of the results we went beyond
the Maxwellian approximation for the gas pdf
and used the so-called first Sonine
correction, which is tantamount to assume
$\phi_S(\mathbf{v})=\frac{m}{2\pi T_g}\exp{ \left( -
\frac{mv^2}{2T_g}\right)}(1+c_2S_2(mv^2/2T_g))$ with $S_2(x)=\frac{1}{2}
x^2- 2 x+1$~\cite{sonine}. In this case one may perform the
calculations retrieving a correction to formula (\ref{eq:temperatura})
of the form
$\eta_S=\eta\frac{8+3c_2}{8-c_2}$, which makes possible
the case $T_r/T_g > 1$, i.e. $\langle V \rangle >0$. Note that in the
literature $\eta$, i.e. the ratio between the gas temperature and the
tracer temperature, has been calculated also for spherical particles
in~\cite{martin}.

Eqs. (\ref{eq:visc})-(\ref{eq:temperatura}) indicate that
a Fluctuation-Dissipation Relation (FDR) holds, 
in contrast with  the 
small violations of FDR reported in studies of different models
of granular tracers ~\cite{fdt}. However, the validity of the FDR
in our case is an effect of the truncation of the KM expansion and of the
small $\epsilon$ approximation considered here.

As anticipated in Figs.~\ref{fig:traj} and~\ref{fig:all}
the validity of the analytical theory has been also tested
against Direct Simulation Monte Carlo (DSMC~\cite{bird}) which enforces the Molecular Chaos assumption, 
used to derive
Eq.~\eqref{eq:boltz}, but in principle  not verified in MD.
In addition  within DSMC it is possible to fix
the desired form  of $\phi(\mathbf{v})$ at our will,
while in MD this depends on the control parameters of the system.
Fig.~\ref{fig:all} displays a good agreement between the theory and 
DSMC results for both observables.
When $\epsilon \sim 1$ the
theoretical results deviate from the DSMC results.
The comparison with MD results is fair, but
as $\alpha_r$ decreases systematic corrections appear.
In particular when $M=1$
we observe a nice agreement with DSMC results, suggesting that in
this case the only source of mismatch with the theory is the high
value of $\epsilon$ rather than the failure of the Molecular Chaos
assumption. At higher values of $\epsilon$
the MD and theoretical results for $M\langle V^2\rangle$ match 
better than the corresponding results for the
average velocity, because the latter originates from a higher order term in the
$\epsilon$ expansion, and therefore is subject to a stronger relative noise.

\begin{figure}[htbp]
\includegraphics[width=7cm,clip=true]{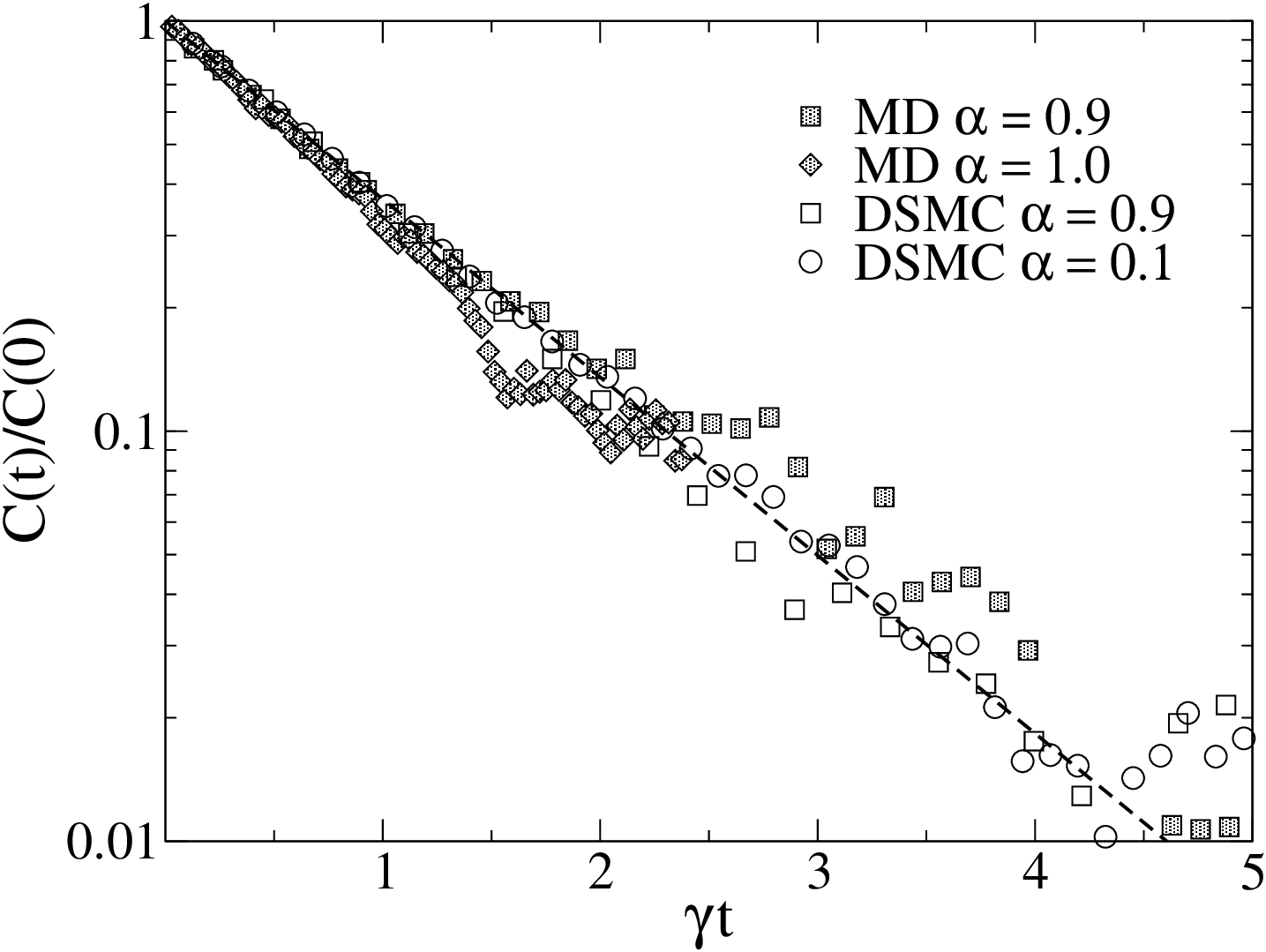}
\caption{Rescaled self-correlation functions of the tracer velocity
with different choices of parameters, elastic and inelastic, from MD
and DSMC, against rescaled time $\gamma t$. The bold dashed line is
the theoretical prediction coming from Eq.~\eqref{eq:visc}
\label{fig:corr}}
\end{figure}

The tracer velocity self-correlation $C(t)=\langle
(V(t)-\langle V \rangle)(V(0)-\langle V \rangle) \rangle$,
displayed in Fig.~\ref{fig:corr} together with the theoretical
prediction $T_r/M \exp(-\gamma t)$, gives a direct measure of $\gamma$
and agrees well with the theory for low inelasticity. For larger
inelasticities, at large time, $C(t)$ shows a fatter tail, which could
be the prelude of some transport coefficient divergence, very well
known for disks in $2D$~\cite{diverge}. In our opinion this is the
explanation for the lack of perfect agreement with the estimates of
$\langle V \rangle$. Nevertheless, the exponential decay of $C(t)$ is consistent with the
observed diffusive behavior at large times, as shown in
Fig.~\ref{fig:traj}. In particular, from Eq.~\eqref{eq:langevin} one
can predict a diffusion coefficient $D=T_r/(M \gamma)$ and a general
formula for the mean squared displacement which reads:
$\frac{d_2(t)\gamma}{2D}=z - [1-\exp(-z)]$ where $z=\gamma t$. This
prediction is fairly confirmed by MD and DSMC simulation, see again
Fig.~\ref{fig:traj}, bottom frame.


We are now in a position to draw conclusions and discuss
perspectives. Our main result is the unveiling of an effect unknown,
to the best of our knowledge, in granular gases: a directed motion
driven by undirected fluctuations, exploiting only the time
irreversibility of inelastic collisions. Experimental verification of
such a phenomenon may easily be achieved: only technical problems
(such as keeping the tracer in contact only with the gas and far from
the external driving, in order to reduce noise, e.g. constraining it
on a suspended guide) must be solved. Predictions can be made for
different ratchets: for instance, we can consider a 
symmetrically shaped tracer, such as a piston or a disk, but with sides (say
left/right) made of different materials, which correspond to different
inelasticities, $\alpha_1$ and $\alpha_2$, respectively. The theory
predicts (at first order in $\epsilon$) a  drift
velocity,  $\langle V \rangle=\sqrt{2\pi m
kT}(u-1)(M/m+\eta-1)/[4M(u+1)]$ with $u=(1+\alpha_1)/(1+\alpha_2)$.
Moreover, other kinds of external drivings can be used: a typical
setup, for example, receives energy from the boundaries. In this case
a box is vibrated and the tracer should be constrained to
be in contact with the gas, but free to move on a $1d$ guide.
Finally, we cannot rule out that similar mechanisms are at work
also in more packed situations and contribute to spontaneous de-mixing
phenomena, such as the Brazil Nut Problem~\cite{bnp}.

\begin{acknowledgments} A. P. acknowledges the Marie Curie grant No. 
MERG-021847 and  UMBM acknowledges a grant COFIN-MIUR 2005, 2005027808.

\end{acknowledgments}



\begin{thebibliography}{99}

\bibitem{intro0} H.M.~Jaeger, S.R.~Nagel and R.P.~Behringer,
Rev. Mod. Phys., 68, 1259 (1996) and references therein.

\bibitem{intro1} T.~P\"oschel and S.~Luding (eds.),
{\em Granular Gases}, Lecture Notes in Physics,
Vol. 564, Springer, Berlin (2001).

\bibitem{intro2} A. Puglisi, F. Cecconi and A Vulpiani, J. Phys.:
Condens. Matter {\bf 17}, S2715 (2005) and references therein.

\bibitem{vdb} C. Van den Broeck, R. Kawai, and P. Meurs,
Phys. Rev. Lett. {\bf 93}, 090601 (2004).

\bibitem{rat} M. v. Smoluchowski, Physik. Zeitschr. {\bf 13}, 1069
(1912); R. P. Feynman, R. B. Leighton, and M. Sands, {\em The Feynman
Lectures on Physics I} (Addison-Wesley, Reading, MA, 1963), Chapter
46.

\bibitem{ap} A. Puglisi, P. Visco, E. Trizac and F. van Wijland,
Phys. Rev. E {\bf 73}, 021301 (2006).

\bibitem{kicks_noi} A. Puglisi, V. Loreto, U. M. B. Marconi, A.  Petri
  and A.  Vulpiani, Phys. Rev. Lett. {\bf 81}, 3848 (1998);
  A. Puglisi, V. Loreto, U. Marini Bettolo Marconi and A. Vulpiani,
  Phys. Rev. E {\bf 59}, 5582 (1999).

\bibitem{kicks_altri} D. R. M. Williams and F. C. MacKintosh, Phys.
  Rev. E {\bf 54} R9 (1996); T.P.C. van Noije, M.H. Ernst, E. Trizac
  and I. Pagonabarraga, Phys. Rev E {\bf 59}, 4326 (1999).

\bibitem{kicks_exp} J. S. Olafsen and J. S. Urbach, Phys. Rev. E {\bf
60}, 0R2468 (1999).


\bibitem{sonine} S. Chapman and T. G. Cowling, {\em
  The Mathematical Theory of Non-Uniform Gases}, Cambridge U.P., 1970.

\bibitem{martin} P. A. Martin and J. Piasecki, Europhys. Lett. {\bf 46}, 613
  (1999).

\bibitem{fdt} A. Puglisi, A. Baldassarri and V. Loreto, Phys. Rev. E
{\bf 66}, 061305 (2002); A. Barrat, V. Loreto and A. Puglisi, Physica
A {\bf 334}, 513 (2004); V. Garz\'o, Physica A {\bf 343}, 105 (2004).

\bibitem{bird} G. A. Bird, Molecular Gas Dynamics and the Direct Simulation of
  Gas Flows, Clarendon 1994 (Oxford); J.M. Montanero and A. Santos, Granular
  Matter {\bf 2}, 53 (2000).

\bibitem{diverge} B.J. Alder and T.E. Wainwright, Phys. Rev.  A {\bf 1}, 18
(1970); M.H. Ernst, E.H. Hauge and J.M.J. van Leeuwen,
Phys. Rev. Lett. {\bf 25}, 1254 (1970); J.R. Dorfman and E.G.D. Cohen,
Phys. Rev. Lett. {\bf 25}, 1257 (1970).

\bibitem{bnp} A. Rosato, K. J. Strandburg, F. Prinz, and R. H. Swendsen,
Phys. Rev. Lett. {\bf 58}, 1038 (1987).



\end{thebibliography}
\end{document}